# Formation of negative hydrogen ions in 7-keV OH$^+$ + Ar and OH$^+$ + acetone collisions: a general process for H-bearing molecular species

Zoltán Juhász[1]*, Béla Sulik[1], Jimmy Rangama[2], Erika Bene[1], Burcu Sorgunlu-Frankland[2], François Frémont[2] and Jean-Yves Chesnel[2]

[1]*Institute for Nuclear Research, Hungarian Academy of Sciences (MTA Atomki), H-4001 Debrecen, P.O. Box 51. Hungary.*

[2]*Centre de Recherche sur les Ions, les Matériaux et la Photonique (CIMAP), Unité Mixte CEA-CNRS-EnsiCaen-Université de Caen Basse-Normandie, UMR 6252, 6 Boulevard Maréchal Juin, F-14000 Caen, France.*

*Correspondence to: zjuhasz@atomki.hu.

Receipt date:

**Abstract**: We demonstrate that the formation of negative hydrogen ions (H$^-$) occurs in a wide class of atomic and molecular collisions. In our experiments, H$^-$ emission from hydroxyl cations and acetone molecules was observed in keV-energy collisions. We show that hydride (H$^-$) anions are formed via direct collisional fragmentation of molecules, followed by electron grabbing by fast hydrogen fragments. Such general mechanism in hydrogen-containing molecules may significantly influence reaction networks in planetary atmospheres and astrophysical media and new reaction pathways may have to be added in radiolysis studies.

The formation of negative ions has been a subject of great interest over the past decades [1-4]. Anions play a major role in many areas of physics and chemistry involving weakly ionized gases and plasmas [1,2,5]. Even in small concentration, anions influence appreciably the properties of their environment [1,6,7].

Since its prediction as a bound system [8], the hydride H$^-$ ion has drawn considerable attention because of its involvement in many reactions important for cosmology [2] and astrophysics [3,9,10]. In the early stage of star formation, H$^-$ ions are involved in the formation of H$_2$ through associative detachment of H$^-$ and H species [2,10]. Additionally, it is commonly accepted that the main source of opacity in the solar atmosphere at red and infrared wavelengths is light absorption by H$^-$ ions [3,6,9]. The presence of H$^-$ is also predicted in various regions of the interstellar medium [7], and in the transition zones of planetary nebulae [10]. Because of the high reactivity of H$^-$, a variety of collisional and chemical processes are expected in zones where both molecules and H⁻ anions are present [11]. Likewise, H⁻ is involved in the chemistry of planetary atmospheres [5]. Even in the case of small yields, H⁻ formation mechanisms are of prime interest.

In the gas phase, slow electron impact on H-containing molecules may lead to H⁻ formation [12,13]. Hydrogen anions can also be formed in collisions between positive ions and neutral atoms or molecules. Electron transfer in keV proton-atom and proton-molecule collisions leads to H⁻ emission [14,15]. In addition to H⁺ impact, collisions of homonuclear hydrogen ions ($H_2^+$ and $H_3^+$) with neutral species may result in the formation of H⁻ anions [16-21]. In the case of keV $H_2^+$ projectiles, H⁻ production was attributed to dissociative electron capture [17,18], e.g., $H_2^+ + Ar \rightarrow (H_2^-)^* + Ar^{2+} \rightarrow H^- + H + Ar^{2+}$. For a few keV $H_3^+$ + He collisions, it was concluded that the H⁻ fragments were formed by the electronic excitation of $H_3^+$ followed by its fragmentation $[(H_3^+)^* \rightarrow H^+ + H^- + H^+]$ [20,21]. All the identified processes [16-21], so far, involve specific excited states of molecular species, which dissociate to H⁻. Momentum exchange between the colliding partners was assumed to be negligible in these processes [16-21].

In this Letter, we show that H⁻ anions can be formed in molecular collisions under much more general conditions via a previously unrecognized process. We demonstrate that H⁻ formation by collision-induced molecular fragmentation can originate from a wider variety of heteronuclear molecular species, such as organic compounds. While investigating collisions at energies relevant to the solar wind and low-temperature astrophysical and laboratory plasmas, we observed H⁻ formation in collisions of a simple molecular cation, OH⁺ incident on Ar atoms and acetone molecules ($CH_3$-CO-$CH_3$). The hydroxyl cation OH⁺ is a key partner in many collisional and chemical processes in interstellar media [22] and planetary atmospheres [23]. Here, we report on the discovery of the process of H⁻ formation via quasi-elastic two-body collisions involving a large momentum exchange between a heavy atomic center and the active H center.

The experiment was performed at the ARIBE facility of the Grand Accélérateur National d'Ions Lourds (GANIL) in Caen, France. The experimental setup was similar to the one depicted in [24]. A molecular OH⁺ beam of ~10 nA was collimated to a diameter of ~2.5 mm before entering the collision chamber. In the center of the chamber, the OH⁺ ions crossed an effusive gas jet of either argon atoms or acetone molecules. In the collision area, the density of the gas target was typically ~$10^{12}$ cm⁻³ at chamber pressures below $1.5 \times 10^{-5}$ mbar. This is the regime of single collisions, where a projectile ion has practically no chance for two independent collisions within the target region. The background pressure without target gas was lower than $3 \times 10^{-7}$ mbar. The electrons and negative ions produced in the collision were detected by a channel electron multiplier of a single-stage electrostatic spectrometer. In the energy dispersive analyzer of the spectrometer, the trajectory of a charged particle is determined by its energy per charge ratio. The spectrometer was mounted on a movable ring, allowing the detection at angles ranging from 30° to 150° with respect to the beam direction.

Figures 1a and 1b show experimental differential cross sections with respect to angle and energy for the emission of negatively charged particles in 7-keV OH⁺ + Ar and OH⁺ + acetone collisions, respectively. In such slow collisions, ionization processes result in a broad distribution of electrons emitted predominantly at low energies. Electron emission is not the subject here; we only mention that in OH⁺ + Ar collisions, a peak structure due to MNN Auger electron emission from argon is observed in the range 10-20 eV (Fig. 1a). Here, the most remarkable feature is the appearance of well-defined peaks above 200 eV which cannot be attributed to electron emission. The intensity of these peaks was found to be proportional to the target density, providing evi-dence that the overwhelming majority of the detected particles leading to these peaks originate from single collisions.

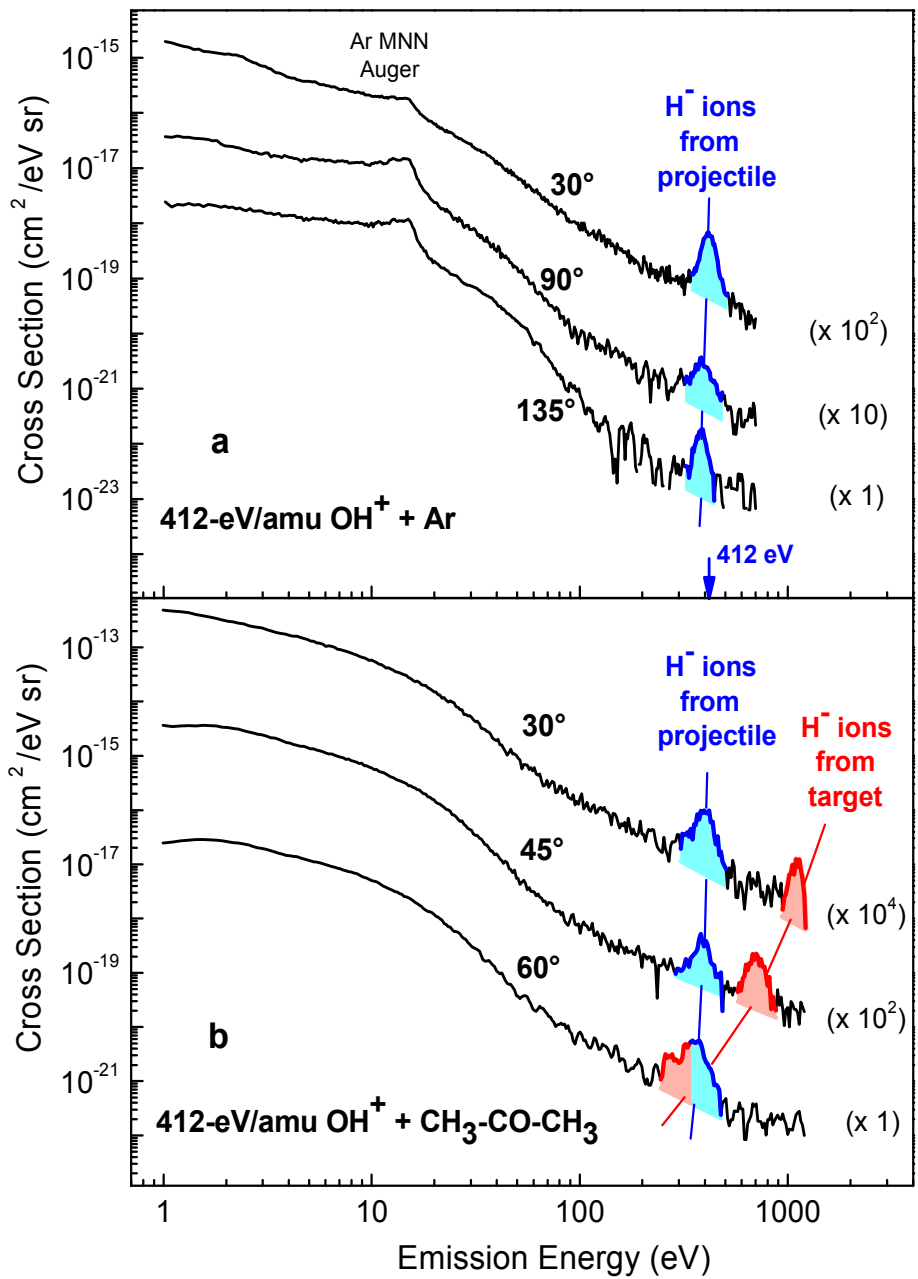

FIG. 1. (Color online) Cross sections for emission of negatively charged particles at different observation angles. (a) 7-keV $OH^+$ + Ar collisions. (b) 7-keV $OH^+$ + acetone collisions. The curves represent electron emission, while the color-shaded peaks represent $H^-$ emission. $H^-$ ions emitted from $OH^+$ projectiles are observed in the peak at about 300-400 eV.
$H^-$ emission from the acetone target results in a second peak, which appears at ~ 1100 eV at 30° and merges with the other peak at 60°. The multiplication factors on the right side are used only for graphical reasons.

For both argon and acetone targets a pronounced peak is observed in the range 300-400 eV at all observation angles. At forward angles (smaller than 90° with respect to the beam direction) the peak energy is close to the kinetic energy per nucleon of the projectile (412 eV) and decreases monotonically with increasing angle. Based on a simple kinematic calculation that includes the target recoil effect (see relations 1a and 1b of Ref. 25) we could assign this peak as to be due to the emission of H$^-$ ions moving with nearly the velocity of the OH$^+$ projectile (Fig. 2a). The presently observed H$^-$ ions are not only removed from the projectile, but they are also scattered to large angles in a hard (large momentum transfer) collision with a heavier target atom (Ar) or atomic core (C or O). In this *two-body* collision, the H nucleus approaches close enough to the heavier nucleus to be effectively deflected by the Coulomb force. As the H nucleus leaves the collision partners due to the Coulomb scattering it grabs two electrons from the collision complex. The inelastic energy transfer in these binary collisions was determined by fitting the angle-energy relation for the scattered center (relation 1b of Ref. 25) to the measured energy of the H$^-$ ions (Fig. 2a). It was found to be close to zero within the limits of ±5 eV, close to the experimental uncertainties.

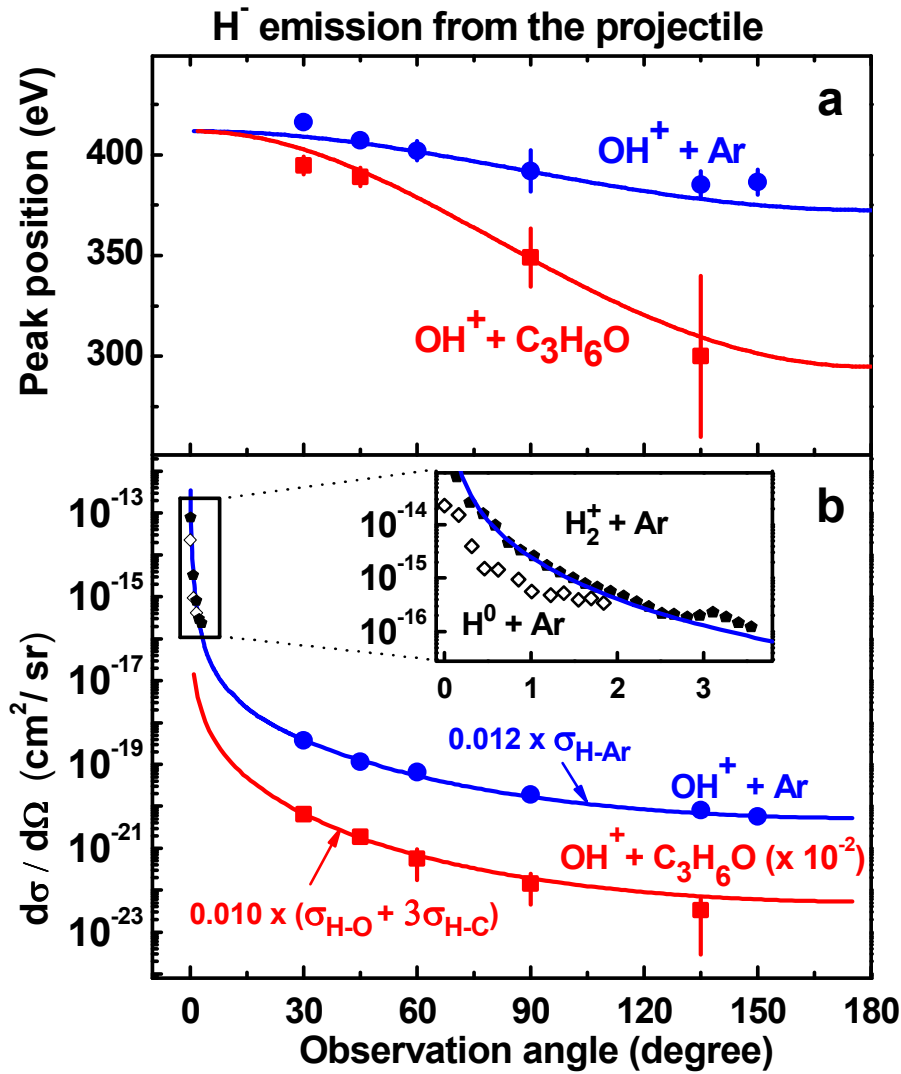

FIG. 2. (Color online) Peak energy and emission cross section for H$^-$ ions ejected from the OH$^+$ projectile.

(a) Measured mean kinetic energy of the H⁻ ions (symbols) and expected energy of H centers elastically scattered on heavy target atoms (curves). For acetone, a weighted average of the results for the C and O atoms is presented.

(b) Present work: Measured H⁻ emission cross section (for Ar (●) and acetone (■) targets) and two-body elastic scattering cross sections multiplied by constant factors (curves). For acetone the curve represents the sum for the contributions of the C and O centers. All data for acetone are multiplied by $10^{-2}$ for graphical reasons. Other works: Measured H⁻ emission cross sections on Ar target for 1-keV $H_2^+$ (◆) [18] and $H^0$ (◇) [26] projectiles.

To our best knowledge, similar H⁻ formation processes were not observed previously. In Double Charge-Transfer spectroscopy [14] and in molecular dissociation experiments [16-21], H⁻ formation was studied only at small scattering angles, at which the momentum transfer between atomic cores is small. In contrast, we found that a hard two-body collision can result in the emission of a fast negatively charged fragment from a positively charged net collision complex. At first sight, formation of fragile H⁻ atomic systems in hard two-body collisions is somewhat surprising, and has not been considered earlier. This process can be referred to as electron grabbing by the fast H fragment.

Another important feature of the $OH^+$ + acetone collision is the existence of a second sharp peak at higher energies (Fig. 1b). This is a recoil peak, observed only at forward angles. While it appears at about 1100 eV at 30°, its energy decreases strongly with increasing observation angle. At 60°, it merges with the other peak around 350 eV and disappears at backward angles (larger than 90°). The kinematic analysis (Fig. 3a) shows that this peak is due to H⁻ emission from the target in hard, quasi-elastic two-body collisions between a target H center and the projectile O center.

Beyond the level of a kinematic analysis, singly-differential cross sections for H⁻ emission from the projectile are presented in Fig. 2b as a function of the observation angle. The cross sections were determined by Gaussian fitting after subtracting the background of the continuous spectra and by integration of the peaks over the emission energy. Strong forward enhancement is visible for both collision systems. Since the energy of the fragments is satisfactorily described in terms of two-center collisions with negligible inelasticity, we compared the measured cross sections with a classical two-body potential-scattering calculation. The scattering potential has been determined by calculating the relaxed ground state energy of the collisional diatomic molecules H-X (X: Ar, C or O) as a function of the internuclear distance $R$ (with $E=0$ at $R→∞$). Here, the collisional molecule H-X is a transient quasi-molecule formed by an H center of one collision partner and by a heavier atomic center X of the other partner. The energy curves were computed by means of the Molpro quantum chemistry software package [27]. Since the present calculations were performed for the ground state of the relaxed diatomic molecule, the obtained interaction potential represents a fully adiabatic molecular situation. This is a reasonable approximation here, because the projectile velocity is about ten times smaller than the mean velocity of the electrons, and thus, the electron orbitals evolve in a quasi-adiabatic manner.

The angular dependence of the calculated results follows that of the measured cross sections within the experimental uncertainties (Fig. 2b). The experimental data fit fairly well to the calculated cross sections multiplied by a constant factor. Thus, the cross section for producing H⁻ ions in hard quasi-elastic collisions is proportional to the cross section for two-body scattering of the H component of the projectile, *i.e.*, the fraction of anions among the emitted H centers does not depend on the emission angle. The H⁻ fraction is found to be comparable for both systems: it is 1.2 % and 1.0 % in case of argon and acetone targets, respectively.

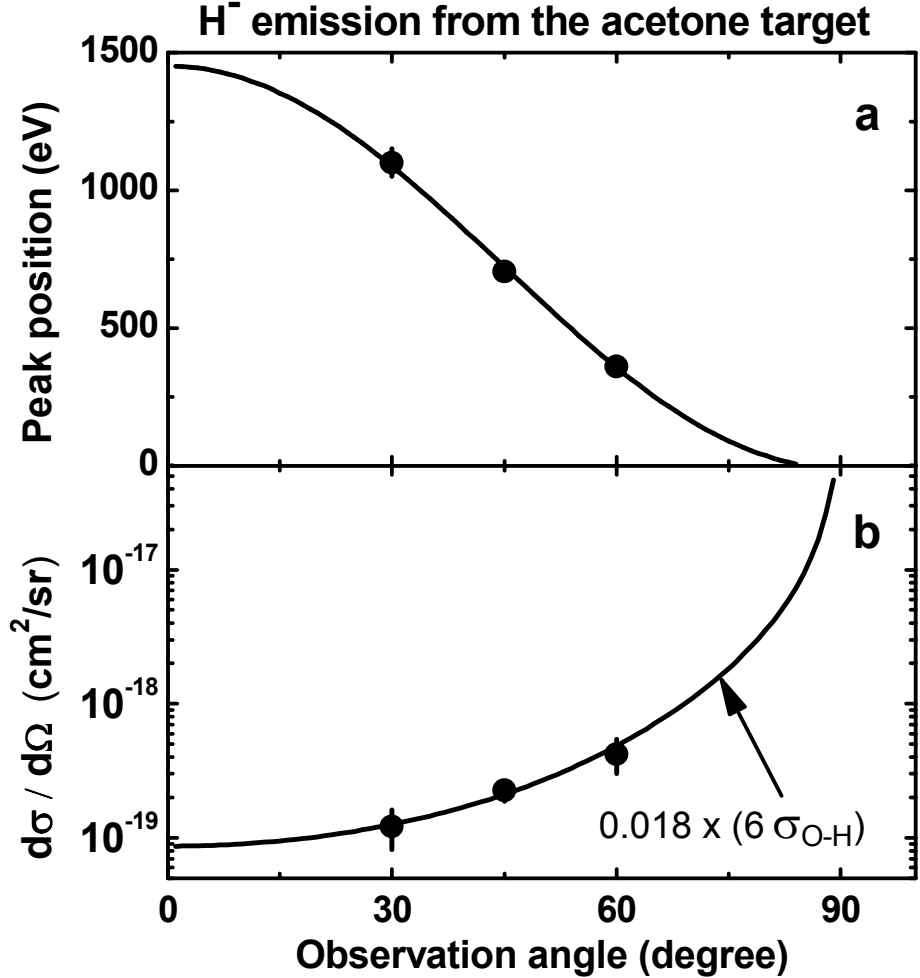

FIG. 3. Peak energy and emission cross section for the H⁻ ions ejected from the acetone target.

(a) Measured mean kinetic energy of the H⁻ ions (symbols) and expected energy of the H center after elastic two-center (O-H) collisions (curve).

(b) Measured cross sections for H⁻ emission (symbols) and two-body elastic scattering cross section summed over the six H atoms of acetone and multiplied by 0.018 (curve).

In the evaluation process, the best estimated values were taken for all the parameters (target density, detector efficiency, etc.). The largest systematic uncertainty emerges from the target density determination at the collision center. It was deduced from comparison of signal strengths for concentrated (when the gas nozzle is close to the collision center) and evenly distributed gas target (the nozzle is far). The corresponding pressure reading (far from the center) is also subject to different gage sensitivity factors for different gases. For our Bayard-Alpert ionization gage we used 1.3 for Ar and 3.6 for acetone. The resulting systematic uncertainty for cross sections is estimated to be about a factor of 2. The relative uncertainty of the measured H⁻ production cross sections (for a given gas target) is generally much smaller. For argon it is smaller than 30%. It was estimated from the statistical uncertainties, and checked by the reproducibility of independent measurements.

Though we could not measure near zero degrees with our apparatus, from the validity of the two-body picture we expect significantly higher yields at small angles for H$^-$ ions emerging from the projectile (Fig. 2b). This expectation is also supported by the fact that at small scattering angles (below 3°) our calculated cross sections follow closely the previously measured cross sections for H$^-$ emission from H$_2^+$ projectiles impacting the same target (Ar) at about the same velocity as the present one (inset in Fig. 2b) [18]. The results for atomic projectiles H$^0$ are not far off either [26]. This good matching of our calculation results with both the earlier and the present experimental data suggests that the H$^-$ emission yield is proportional to the two-body collision cross section in a wide range of scattering angles even down to fractions of a degree.

Similar results are shown in Fig. 3b for H$^-$ emission from the acetone target. The measured H$^-$-production cross section and the calculated two-body (O-H) scattering cross section for the recoiling H center are found to be proportional. This shows that the H$^-$ fraction of the recoiling H centers is angle independent. From the proportionality factor, the H$^-$ fraction is found to be ~1.8%. One should note that the cross section for H$^-$ emission from the target must reach a maximum near 90° (Fig. 3b). But, kinetic energies of H$^-$ ions emitted at angles close to 90° do not exceed a few eV (Fig. 3a). Due to the emission of low-energy electrons in large yields (Fig. 1), the contribution of slow H$^-$ ions to the recorded spectra cannot be determined here.

We note that hydride anions may appear not only in the two-body peaks, but also in a wide range of emission energies. Sufficient energy and momentum transfer to liberate H centers can occur in three- or many-body collisions. Moreover, in softer collisions, the earlier observed mechanisms involving excited repulsive states [15-21] may also contribute to H$^-$ production from any H-containing molecules. Accordingly, the total H$^-$ yield is likely to be larger than those determined from the peaks observed in Fig. 1. We conclude that H$^-$ formation is a general process in collisions of hydrogen-containing molecular species at keV impact energies.

A remarkable feature of the present process is the fact that – at a given collision velocity – the relative H$^-$ yield appears to be independent of the observation angle and only slightly depends on the parent molecule of the ejected H center (Figs. 2b and 3b). This finding indicates the generality of the discovered process of two-electron-grabbing by an H fragment formed in a hard two-body collision. Moreover, this finding provides new insight into the physics of the collision. It shows that for a given collision system the H$^-$ fraction remains constant with changing the momentum transfer between atomic cores. According to the present calculations, the distance of closest approach between the atomic centers involved in the present hard two-body collisions is of the order of 1 atomic unit (a.u.). In such close collisions, the electronic clouds of the collision partners merge each other so that the collision complex forms a quasi-molecule in which most of the electrons are shared. Hence, before escaping the collision complex, the receding proton moves through a reservoir of electrons from which it grabs zero, one or two electrons. Here, the collision velocity is small enough to allow a nearly adiabatic development of the electronic states. The independence of the H$^-$ fraction from the momentum transfer suggests the scenario that many capture channels are open to end up in a limited number of final states and result in a statistical distribution of the final charge state (0, +, or −) of the emitted H centers.

For estimating the energy dependence of the discovered process, one should consider its two main steps. First a proton should be liberated, which needs a minimum energy transfer ($\Delta E_{min}$ ~ 10-20 eV). The cross section for sufficient energy transfer ($\Delta E > \Delta E_{min}$) decreases with the impact energy $E$ (for $E \gg \Delta E_{min}$, $\sigma^{Rutherford} \sim 1/E$). The tendency is opposite for electron transfer. Cross section for double electron transfer to the proton increases with increasing $E$ up to ~5 keV/u and exhibits a plateau up to 40 keV/u in proton-Ar collisions [15]. Similar energy dependence has been found for the single capture to a hydrogen atom resulting in H⁻ [26]. Accordingly, the studied process is likely to show a maximum cross section in the 1-5 keV/u collision energy range. Further experiments are planned for mapping its significance in a wide range of impact energy.

In conclusion, we found pronounced peaks superimposed on the electron emission spectra following keV molecular collisions, which are due to H⁻ emission from both the projectile ($OH^+$) and the H-containing molecular (acetone) target. The angular dependence of the mean kinetic energy of the H⁻ peaks can be accounted for by the simple kinematics of a nearly elastic two-body collision between an H center of one collision partner and a heavier center of the other partner. Moreover, the cross section for H⁻ production is proportional to that of the classical two-body potential scattering of the corresponding two centers. From these results, our main findings are the following: (i) fragile atomic systems such as hydride anions are formed not only in soft collisions involving negligible momentum transfer, but also in hard core-core collisions via double electron grabbing by fast hydrogen fragments, (ii) H⁻ ions can originate from any H-containing molecules, (iii) for a given multi-electron collision system the H⁻ fraction among the emitted H centers is independent of the momentum transfer, suggesting the ejection of H centers with a statistical distribution of their final charge state.

As a brief comment on the possible impact of the present findings, we note that the existence of the process reported here is a starting point to study its details and its significance in different fields. In heavy ion therapy and radiolysis, besides aggressive radicals, hydride anions may also be produced from water and organic molecules in regions where the ionic projectiles slow down to a few keV in the tissue and reach their equilibrium (low) charge state. In dense media, formation of H⁻ may be further enhanced by the fact that each projectile suffers several collisions before being stopped. The question arises as to whether these H⁻ ions can deactivate some of the aggressive radicals or initiate hitherto unexpected chemical reactions. Furthermore, our finding of H⁻ emission from isolated molecules colliding at keV energies is of prime relevance for any low density medium, where H-containing molecules are present and collisions take place at solar wind velocities. We thus anticipate that the formation of highly reactive H⁻ anions in a large variety of collisions may influence the collision and reaction networks in astrophysical media and planetary atmospheres.

**Acknowledgments:** This work was supported by the Transnational Access ITS-LEIF, the European Project HPRI-CT-2005-026015, the Hungarian National Science Foundation OTKA (K73703) and the French-Hungarian Cooperation Program PHC Balaton (N° 27860ZL / TÉT_11-2-2012-0028). Thanks to N. Stolterfoht, J.A. Tanis and R.G. Lovas for critical reading of the manuscript and to L. Maunoury and F. Noury for high-level technical assistance.